\documentclass{article}

\usepackage{array}
\usepackage[sorting=none, style=phys]{biblatex}
\usepackage{authblk}
\usepackage{float}
\usepackage{amsmath}
\usepackage{amsfonts} 
\usepackage{physics,comment,cancel}
\usepackage[colorlinks=true, citecolor=cyan, linkcolor=blue, urlcolor=red]{hyperref}
\usepackage[margin=64pt]{geometry}
\usepackage{tikz}
\usepackage{amsthm}
\usepackage[toc,page]{appendix}
\usepackage{dsfont}
\usepackage[table]{xcolor}

\newtheorem{lemma}{Lemma}

\usetikzlibrary{perspective,3d}

\newcommand{\red}[1]{{\color{red} #1}}

\newcommand{\com}[2]{\left[#1,#2\right]}
\newcommand{\acom}[2]{\left\{#1,#2\right\}}

\title{The Mermin-Peres Magic Square Game With Three or More Players}
\author[1]{Itamar Lorian}
\author[1,2]{J.E. Avron}
\affil[1]{Department of Physics, Technion, Israel Institute of Technology, Haifa, 3200, Israel}
\affil[2]{The HDQC, Technion, Haifa, 3200, Israel}
\date{May 2026}

\begin{document}

\maketitle

\begin{abstract}
We present a generalization of the Mermin-Peres magic square game to \(N\) players, in which the square is replaced by an \(N\)-dimensional hypercube. We show that the winning probability of the optimal classical strategy remains \(8/9\) for all \(N\), while for \(N\geq3\), a generalization of the quantum strategy for \(N=2\) achieves a winning probability between \(1/3\) and \(3/4\). This loss of quantum advantage for \(N\geq3\) is a consequence of monogamy.
\end{abstract}

\section*{Author contributions}

All authors contributed equally to this work. During the preparation of this manuscript we used large language models (LLMs) to improve phrasing, check grammar and spelling, ensure consistency of notation, and help verify and simplify a few proofs. An LLM also suggested the lower bound given in \eqref{eq:qlb}, although the proof we give is better.

\section{Introduction}

The Mermin-Peres magic square game offers an example of the strict advantage that quantum resources can give. The magic square was originally introduced by Cabello \cite{Cabello} based on the works of Mermin \cite{mermin, MerminAJP} and Peres \cite{AP}, and later popularized by Aravind \cite{aravind2002, aravind2004} and Brassard et al \cite{Brassard}.

\medskip

The game features two players, Alice and Bob, who play together against a referee. In order to win, Alice and Bob may agree on a strategy and make any preparations they wish before the start of the game. However, once the game begins, they are no longer allowed to communicate with one another. Utilizing these restrictions, the game is constructed in such a way that if Alice and Bob have no access to quantum resources, their strategy cannot guarantee victory. However, if they can make perfect quantum preparations and measurements, they can devise a strategy which will always win. A review of other games that showcase this particular advantage of quantum resources is given in \cite{Brassard}. 

\medskip

Here we study a particular generalization of the game to \(N\) players, as it is possible to formulate more than one seemingly natural generalization. The generalization replaces the magic square with a magic \(N\)-dimensional hypercube, and adjusts the rules of the game appropriately.

\medskip

We find that for players who use an optimal classical strategy, the winning probability stays \(8/9\). Meanwhile, for three or more players who pursue a generalization of the quantum strategy used for the original magic square game, the optimal winning probability satisfies 
\begin{equation}
\frac13\le\text{Prob\((\)Win\()\)}\le\frac34,
\end{equation}
This result is an expression of monogamy \cite{vidick, Polygamy}, and it shows that entanglement need not guarantee quantum advantage. 

\section{The magic square game}

\subsection{The rules of the game}\label{sub:rules2}

We start with a review of the two-player magic square game, in a way that prepares the ground for the generalization to \(N\) players. The rules of the game can be summarized as follows:
\begin{itemize}
\item The referee presents the two players (Alice and Bob) with a \(3\times3\) table.
\item At the start of the game, the referee covertly selects a cell in the table uniformly at random, and assigns its row to Alice and its column to Bob\footnote{This is equivalent to the referee choosing a random row and a random column with uniform probabilities, which is the way the game is often presented. The motivation for this subtle change is explained in \autoref{sub:rulesN}}. Each player is ignorant of the row or column assigned to the other player.
\item  Alice needs to assign each cell in her row either \(1\) or \(-1\), such that the product of the three cells is equal to \(1\). Similarly, Bob needs to assign each cell in his column either \(1\) or \(-1\), such that the product of the three cells is equal to \(-1\).
\item The players win if, at the cell selected by the referee, they both assign the same value.
\end{itemize}
As mentioned in the introduction, Alice and Bob are free to coordinate a strategy and make any preparations necessary before the game begins. Once the game starts, however, they can no longer exchange information.

\subsection{Classical strategy}

When Alice and Bob are restricted to classical means, they cannot coordinate a strategy that guarantees victory. Since each cell in the table is equally likely to be selected as the common cell, a winning strategy would require the players to agree upon a shared table before the start of the game. This table must satisfy the game's rules, meaning that the product of the cells in each row must be \(1\), while the product of the cells in each column must be \(-1\). However, no such table exists. That is because, taking the product over all the rows would yield \(1\), whereas taking the product over all the columns would yield \(-1\). But since both computations correspond to taking the product over all the cells in the table, this results in a contradiction. Therefore, the best Alice and Bob can do is agree upon a table that guarantees winning for eight of the nine cells, like the table shown in \autoref{fig:classic2}. Hence, since the common cell is selected with equal probability, the probability that Alice and Bob will win is at most \(8/9\). 

\begin{figure}[H]

\centering
\input{classic2.tex}
\caption{An example of a $3\times 3$ table in which Alice's and Bob's assignments differ in only one cell, indicated by \(\pm 1\). To satisfy the game's rules, Alice must assign \(1\) to this cell while Bob must assign \(-1\). The red column on the right marks Alice's row constraints, and the red row on the bottom marks Bob's column constraint.}

\label{fig:classic2}
\end{figure}

\subsection{Quantum strategy}\label{sub:quantum2}

Using quantum mechanics however, Alice and Bob can devise a strategy that guarantees victory. To achieve this, the first thing Alice and Bob need to do is agree upon a \(3\times3\) table where each cell is assigned an observable constructed from a tensor product of two Pauli operators (including the identity) which abides by the following constraints:
\begin{itemize}
\item The observables in every row commute and their product is \(\mathds{1}\).
\item The observables in every column commute and their product is \(-\mathds{1}\). 
\end{itemize}
A simple algorithm for constructing such a table is given in Appendix \ref{a:demagic}, and an example of one is shown in \autoref{fig:quantum2}.

\begin{figure}[H]

\centering
\input{quantum2.tex}
\caption{An example of the \(3\times3\) table Alice and Bob need to agree upon in the quantum strategy. The red column on the right marks the constraints on the rows' products, and the red row on the bottom marks the constraints on the columns' products.
}

\label{fig:quantum2}
\end{figure}

\medskip

After Alice and Bob agree on a table that abides by these constraints, they need to prepare a specific \(4\)-qubit quantum state and divide it between them. Given this elaborate setup, once the game begins all Alice and Bob need to do is to measure the observables corresponding to the row or column assigned to them by the referee. Since the observables they measure are tensor products of Pauli operators, the result for each measurement would be either \(1\) or \(-1\), which they then need to assign to the matching cell in the table the referee presented them with.

\medskip

Reasoning why the product of the measurements in any given row is necessarily \(1\), while the product of the measurements in any given column is necessarily \(-1\), as required by the rules of the game is simple enough. Although the values Alice measures for a given row are random, their product is guaranteed\footnote{This follows from the fact that the product of commuting operators reduces to the product of measurement outcomes, which can be seen by looking at the identity in the basis where the operators are diagonal.} to be \(1\). Similarly, the product of the values Bob measures for a given column is guaranteed to be \(-1\). What remains, then, is to determine how Alice and Bob can ensure that they assign the same value to the common cell selected by the referee.

\medskip

To see how, let \(\mathbf{P}\) denote the observable assigned to the common cell the referee has selected, and let \(\ket\psi_{AB}\) denote the shared \(4\)-qubit state prepared by Alice and Bob. Given this notation, we can determine that Alice and Bob will always assign the same value to the common cell, provided the state they prepare satisfies
\begin{equation}\label{eq:CC}
\mathbf{P}\otimes\mathbf{P}\ket\psi_{AB} = \ket\psi_{AB}.
\end{equation}
Since any one of the cells in the table can be selected to be the common cell, this equation must hold for all nine of them. Meaning, the state Alice and Bob need to prepare must be a stabilizer of the Pauli group generated by the observables \(\mathbf{P}\otimes\mathbf{P}\). While it is not immediately clear, such a nontrivial state exists for any table satisfying the constraints we mentioned, and is generally given by
\begin{equation}
\ket\psi_{AB}=\ket{\phi_1}_{AB}\otimes\ket{\phi_2}_{AB},\quad\ket{\phi_1}_{AB},\ket{\phi_2}_{AB}\in\mathcal{B},
\end{equation}
where \(\mathcal{B}\) is some subset of the Bell states which depends on the exact table chosen. For the table given in \autoref{fig:quantum2}, we obtain that
\begin{equation}
\mathcal{B}=\left\{\ket{\Psi^-}_{AB}\right\},\quad\ket{\Psi^-}_{AB}=\frac{\ket{0}_A\otimes \ket{1}_B-\ket{1}_A\otimes \ket{0}_B}{\sqrt 2}.
\end{equation}
Meaning,
\begin{equation}\label{eq:ES}
\ket{\psi}_{AB} = \ket{\Psi^-}_{AB}\otimes \ket{\Psi^-}_{AB}.
\end{equation}
This is all rigorously shown in \autoref{a:singlet}.

\medskip

\section{The \textit{N}-dimensional hypercube game}

\subsection{Preliminaries}\label{sub:pre}

We now turn to the generalization of the magic square game to \(N\) players. But before discussing the game itself, it will be useful to establish some notation.

\medskip

Since the \(N\)-player version of the game will involve an \(N\)-dimensional $3^{\otimes N}$ table, we shall use \(C\) to denote the cells of this table, where each cell is labeled by a collection of \(N\) indices
\begin{equation}
C(i_1,\dots,i_N), \quad i_1,\dots,i_N\in\{1,2,3\}.
\end{equation}
Next, we need to define the notion of a row (column) in \(N\) dimensions. There are two natural possibilities. The first is to define a row as a one-dimensional slice of the table parallel to one of its axes, and the second is to define a row as an \((N-1)\)-dimensional slice of the table orthogonal to one of its axes. Each choice will result in a different generalization of the game, and subsequently different strategies. In this paper we will focus only on the former choice. Therefore, we shall use \(L_n\) to denote the rows which are parallel to the \(n\)-th axis of the table, where each row is labeled by a collection \(N-1\) indices
\begin{equation}\label{eq:columns}
 L_n({i_1,\dots,\cancel{i_n},\dots, i_N})
 =\bigcup_{i_n\in\{1,2,3\}}C({i_1,\dots,i_n,\dots,i_N}).
\end{equation}
Finally, it will also be helpful to use \(S_{m,n}\) (with \(m<n\)) to denote the two-dimensional slices which are parallel to the \(m\)-th and \(n\)-th axes of the table, where each slice is labeled by a collection of \(N-2\) indices
\begin{equation}\label{eq:S}
S_{m,n}(i_1,\dots,\cancel{i_m},\dots,\cancel{i_n},\dots,i_N)
=\bigcup_{i_m,i_n\in\{1,2,3\}}C({i_1,\dots,i_m,\dots,i_n,\dots,i_N}).
\end{equation}

\subsection{The rules of the game}\label{sub:rulesN}

Looking at the rules of the original game as stated in \autoref{sub:rules2}, and using the notion of a row in \(N\) dimensions introduced in \autoref{sub:pre}, the natural generalization to \(N\) players is:
\begin{itemize}
\item The referee presents the \(N\) players with an empty \(N\)-dimensional \(3^{\otimes N}\) table.
\item The referee secretly selects a cell in the table uniformly at random, and assigns the \(n\)-th player the corresponding row \(L_n\) which passes through that cell. Each player is ignorant of the rows assigned to the other players.
\item Every player needs to assign each cell in their row either \(1\) or \(-1\), such that the product of the three cells for the \(n\)-th player is equal to \((-1)^n\).
\item The players win if all of them  assign the same value to the common cell.
\end{itemize}
We can now see why it was necessary to reformulate the original rules so that the referee selects a random cell in the table, rather than independent rows (or a row and column). This is simply because in \(N\ge3\) dimensions, \(N\) random orthogonal rows (as we chose to define them) need not intersect in a single cell. In fact, the probability that they do is
\begin{equation}
\text{Prob}\left(\bigcap_{n=1}^N L_n\neq\emptyset\right)= \frac{3^N}{3^{N(N-1)}}=3^{-N(N-2)},
\end{equation}
where the denominator is the total number of orthogonal rows, and the numerator is the number of choices of orthogonal rows which intersect in a single cell. Thus, while for \(N=2\) the two formulations are equivalent, in higher dimensions the original formulation leads to an ill-defined winning condition. Interestingly, we would not have encountered this issue if we instead chose to define a row in \(N\) dimensions as an \((N-1)\)-dimensional slice of the table.

\subsection{Classical strategy}

Similarly to the original game, since each cell in the table is equally likely to be selected as the common cell, a winning strategy requires the players to agree upon a shared table which satisfies the game's rules beforehand. That is, for every row \(L_n\) in the table, the product of its three cells must be \((-1)^n\). However, exactly as in the two-player case, such an assignment is impossible. The problem therefore becomes determining the minimum number of cells on which the players' assignments must disagree.

\medskip

A simple lower bound can be obtained by looking at the two-dimensional slices \(S_{m,n}\), for some \(m-n=1\) (mod \(2\)). Since \(m\) and \(n\) have different parity, in every slice the product over the rows parallel to one axis must equal \(1\), while the product over the rows parallel to the other axis must equal \(-1\). Consequently, each slice is exactly equivalent to the table required for the classical strategy of the two-player game. As we saw when reviewing that strategy, this indicates that there must be at least one cell in any given slice for which the players disagree. From \eqref{eq:S} it follows that there are \(3^{N-2}\) such parallel slices, which means that there must be at least \(3^{N-2}\) cells in which the assigned values of the players differ. Hence, because the table contains \(3^N\) cells, the probability of losing using a classical strategy must be at least
\begin{equation}\label{eq:bound}
\text{Prob\((\)Lose\(\))}\ge\frac{3^{N-2}}{3^N}=\frac19.
\end{equation}
Equivalently, the probability of winning using a classical strategy must be at most \(8/9\).

\medskip

We already saw that this bound is tight for \(N=2\), but is that the case for all \(N\ge2\)? Meaning, can we always construct a classical strategy that saturates the bound? The answer is yes, as we will now show by giving an explicit construction.

\medskip

Let \(T_a\) and \(T_{a,b}\), where \(a,b\in\{1,2,3\}\), denote \(3\times3\) tables in which cell \((i_1,i_2)\) is assigned
\begin{equation}\label{eq:Ta}
T_a(i_1,i_2)=1- 2\delta_{a,i_1} ,\quad T_{a,b}(i_1,i_2)=1-2\delta_{a,i_1}(1-\delta_{i_2, b}),
\end{equation}
where \(i_1,i_2\in\{1,2,3\}\). For example, the tables denoted by \(T_3\) and \(T_{3,2}\) are shown in \autoref{fig:T}.

\begin{figure}[H]

\centering
\input{T.tex}
\caption{The tables denoted by \(T_3\) and \(T_{3,2}\), labeled accordingly.}

\label{fig:T}
\end{figure}

\medskip

We can then formulate a valid classical strategy which saturates the bound given in \eqref{eq:bound}, by having the odd and even numbered players construct two distinct \(N\)-dimensional tables, with the cells of each filled according to the rule
\begin{equation}\label{eq:minus}
 C^{(\text{odd})}(i_1,\dots,i_N)=T_\alpha(i_1,i_2),\quad C^{(\text{even})}(i_1,\dots,i_N)=T_{\alpha,\beta}(i_1,i_2),
\end{equation}
where we denoted
\begin{equation}
\alpha=1\oplus i_3\oplus i_5\oplus\cdots,\quad\beta=1\oplus i_4\oplus i_6\oplus\cdots,
\end{equation}
with \(\oplus\) signifying addition modulo 3 on the  set \(\{1,2,3\}\).

\medskip

To see why this is a valid strategy, we must show that it satisfies the game's rules. That is, for every player \(n\in\{1,\dots,N\}\), the product of the cells in each row is equal \((-1)^n\). Equivalently, we can show that the parity of the number of cells which are assigned \(-1\) in each row is equal to the parity of \(n\). From \eqref{eq:Ta} and \eqref{eq:minus} it follows that 
\begin{equation}\label{eq:minuscond}
C^{(\text{odd})}(i_1,\dots,i_N)=-1\iff(i_1=\alpha),\quad C^{(\text{even})}(i_1,\dots,i_N)=-1\iff(i_1=\alpha\land i_2\neq\beta)
\end{equation}
Therefore, remembering the definition of a row as given in \eqref{eq:columns}, we obtain that:
\begin{itemize}
    \item For odd \(n\), in each row either \(i_1\) sweeps over \(\{1,2,3\}\) while \(\alpha\) stays fixed (when \(n=1\)), or \(\alpha\) sweeps over \(\{1,2,3\}\) while \(i_1\) stays fixed (when \(n>1\)). Hence, \(i_1=\alpha\) exactly one time, which means that one cell is assigned \(-1\).
    \item  For even \(n\), in each row either \(i_2\) sweeps over \(\{1,2,3\}\) while \(i_1\), \(\alpha\) and \(\beta\) stay fixed (when \(n=2\)), or \(\beta\) sweeps over \(\{1,2,3\}\) while \(i_1\), \(i_2\) and \(\alpha\) stay fixed (when \(n>2\)). Hence, if \(i_1=\alpha\) then \(i_2\neq\beta\) exactly two times, which means that two cells are assigned \(-1\). Otherwise, if \(i_1\neq\alpha\), then zero cells are assigned \(-1\).
\end{itemize}
Overall, we get that the number of cells which are assigned \(-1\) in each row indeed matches the parity of \(n\).

\medskip

Next, to see why this strategy saturates the bound given in \eqref{eq:bound}, we must show that the players disagree in exactly \(3^{N-2}\) cells. From \eqref{eq:Ta} and \eqref{eq:minus} it follows that the even and odd numbered players disagree when \((i_1=\alpha)\wedge(i_2=\beta)\).
Meaning, they disagree in the cells \(C(\alpha,\beta,i_3,\dots,i_N)\), where \(i_3,\dots,i_N\in\{1,2,3\}\). We can thus conclude that the strategy features \(3^{N-2}\) such disagreements, which means that it indeed wins with a probability of \(8/9\).

\subsection{Quantum strategy}

The quantum strategy for the original game consists of two main steps. First, the players construct a \(3\times3\) magic square of observables whose measurement outcomes always satisfy the rules of the game. Second, they prepare and distribute a \(4\)-qubit quantum state that guarantees the same measurement outcome for the common cell. To extend the strategy to \(N\) players, we therefore need to generalize both steps.

\subsubsection{The magic \textit{N}-dimensional hypercube}

The most straightforward generalization of the magic square is a magic \(N\)-dimensional \(3^{\otimes N}\) hypercube, where each cell is assigned an observable constructed from a tensor product of two Pauli operators (including the identity) which abides by the following constraints:
\begin{itemize}
\item The observables in any row \(L_n\) commute and their product is \((\mathds{-1})^n\).
\end{itemize}
By the same reasoning as in the two-player case, these constraints ensure that the product over the measurements for any given row \(L_n\) is \((-1)^n\), as required by the rules of the game. It remains to show that such a table exists for an arbitrary \(N\), which we shall do by giving a recursive construction.

\medskip

We start with the case of \(N=3\). Let \(C(i_1,i_2)\), with \(i_1,i_2\in\{1,2,3\}\), be the cells of some valid two-player table. Since the observables in each row commute, a cyclic permutation of the rows preserves the constraints over the table. Hence, we can construct a valid three-player table by assigning
\begin{equation}\label{eq:recursion}
C(i_1,i_2,i_3)=C(i_1\oplus i_3\ominus 1,i_2)=
\begin{cases}
C(i_1,i_2),&i_3=1\\
C(i_1\oplus1,i_2),&i_3=2\\
C(i_1\oplus2,i_2),&i_3=3
\end{cases},
\end{equation}
with \(\ominus\) signifying subtraction modulo 3 on the  set \(\{1,2,3\}\). The rows parallel to the first and second axes are simply the rows of the two-player table after some cyclic permutation, and therefore preserve the required constraints. Meanwhile, the rows parallel to the third axis consist of
\begin{equation}
L_3(i_1,i_2)=C(i_1,i_2)\cup C(i_1\oplus1,i_2)\cup C(i_1\oplus2,i_2),
\end{equation}
which is precisely a cyclic permutation of row \(L_1(i_2)\) in the two-player table. Thus, the observables in these rows commute and their product is equal to \(\mathds{-1}\), meaning they also preserve the required constraints. Overall, every row in the three-player table satisfies the commutativity and product constraints, so we indeed obtain a valid table. An example of this construction is shown in \autoref{fig:quantum3}, where it is applied to the two-player table presented in \autoref{fig:quantum2}.
\begin{figure}[H]   
\centering
\input{quantum3.tex}
\caption{The three-player table obtained by applying the recursive construction described in \texorpdfstring{\eqref{eq:recursion}}{eq. (recursion)} to the two-player table shown in \texorpdfstring{\autoref{fig:quantum2}}{table 2}.}
\label{fig:quantum3}
\end{figure}

\medskip

Proceeding to the case of \(N=4\), similarly to the previous case, we can construct a valid four-player table by assigning
\begin{equation}\label{eq:recursion2}
C(i_1,i_2,i_3,i_4)=C(i_1,i_2\oplus i_4\ominus 1,i_3)=
\begin{cases}
C(i_1,i_2,i_3),&i_4=1\\
C(i_1,i_2\oplus1,i_3),&i_4=2\\
C(i_1,i_2\oplus2,i_3),&i_4=3
\end{cases}.
\end{equation}
As before, the rows parallel to the first, second and third axes are simply the rows of the three-player table after some cyclic permutation, and therefore preserve the required constraints. Meanwhile, the rows parallel to the fourth axis consist of
\begin{equation}
L_4(i_1,i_2,i_3)=C(i_1,i_2,i_3)\cup C(i_1,i_2\oplus1,i_3)\cup C(i_1,i_2\oplus2,i_3),
\end{equation}
which is precisely a cyclic permutation of row \(L_2(i_1,i_3)\) in the three-player table. Thus the observables in these rows commute and their product is equal to \(\mathds{1}\), meaning they also preserve the required constraints. Hence, we again obtain a valid table.

\medskip

We can continue this construction indefinitely, by assigning the \(N\)-dimensional table
\begin{equation}\label{eq:construction}
C(i_1,\dots,i_N)=C(i_1,\dots,i_n\oplus i_N\ominus 1,\dots,i_{N-1})\ \text{with}\ n\in\{1,\dots,N-1\};\ n=N\ (\text{mod}\ 2).
\end{equation}

\subsubsection{Three  or more players cannot always win the magic hypercube game}

After the players agree on an \(N\)-dimensional table, a generalization of the two-player strategy calls for them to prepare and distribute a \(2N\)-qubit quantum state which ensures that they always measure the same value in the common cell. But as it turns out, when \(N\geq3\) this is impossible. More generally:
\begin{lemma}
Given two anti-commuting Pauli operators \(\mathbf{P}\) and \(\mathbf{Q}\), no state \(\ket{\psi}_{ABC}\) of Alice, Bob and Charlie can guarantee that Alice and Bob will obtain the same result upon measuring \(\mathbf{P}\), while Bob and Charlie obtain the same result upon measuring \(\mathbf{Q}\).
\end{lemma}

To show this, let us denote
\begin{equation}\label{eq:PAB}
    \mathbf{P}_{AB}=\mathbf{P}\otimes\mathbf{P}\otimes\mathds{1},\quad\mathbf{Q}_{BC}=\mathds{1}\otimes\mathbf{Q}\otimes\mathbf{Q}.
\end{equation}
Alice and Bob obtain the same result upon measuring \(\mathbf{P}\) if and only if
\begin{equation}
\frac{\mathds{1}-\mathbf{P}_{AB}} 2 \ket{\psi}_{ABC}=0 \Longleftrightarrow \mathbf{P}_{AB}\ket{\psi}_{ABC}=\ket{\psi}_{ABC},
\end{equation}
while Bob and Charlie obtain the same result upon measuring \(\mathbf{Q}\) if and only if
\begin{equation}\label{eq:QAB}
\frac{\mathds{1}-\mathbf{Q}_{BC}} 2 \ket{\psi}_{ABC}=0 \Longleftrightarrow \mathbf{Q}_{BC}\ket{\psi}_{ABC}=\ket{\psi}_{ABC}.
\end{equation}
Now, since \(\mathbf{P}\) and \(\mathbf{Q}\) anti-commute, so do \(\mathbf{P}_{AB}\) and \(\mathbf{Q}_{BC}\). Therefore, the Pauli group generated by \(\mathbf{P}_{AB}\) and \(\mathbf{Q}_{BC}\) contains anti-commuting elements, which implies the absence of a stabilizer. Explicitly,
\begin{equation}
\ket{\psi}_{ABC}=\mathbf{Q}_{BC} \mathbf{P}_{AB}\ket{\psi}_{ABC}=-\mathbf{P}_{AB} \mathbf{Q}_{BC}\ket{\psi}_{ABC}=-\ket{\psi}_{ABC}.
\end{equation}

\medskip

Because the Mermin-Peres magic square must contain at least one pair of anti-commuting Pauli operators (see \autoref{a:demagic}), it follows that no quantum state can guarantee that \(N\geq3\) players will assign the same value to every cell in the magic hypercube. Hence, no quantum strategy can guarantee a win for three or more players.

\medskip

This raises the question of what winning probability can be achieved by a quantum strategy for \(N\geq3\) players, which we will tackle in the following two subsections.  

\subsubsection{An upper bound on the winning probability for three or more players}\label{s:warmup}

To attain an upper bound on the winning probability of a quantum strategy, let us consider the following question: Given two anti-commuting Pauli operators \(\mathbf{P}\) and \(\mathbf{Q}\), what is the maximum probability for Alice, Bob and Charlie to obtain consensus upon measuring either \(\mathbf{P}\) or \(\mathbf{Q}\), when chosen uniformly at random?

\medskip

Since $\mathbf{P}$ and $\mathbf{Q}$ are assigned with equal probability, the probability for Alice, Bob and Charlie to obtain consensus in the state \(\ket{\psi}\) is given by
 \begin{equation}\label{eq:probcon}
     \text{Prob(Consensus)}=\frac12\bra\psi\Pi_\mathbf{P}\ket\psi+\frac12\bra\psi\Pi_\mathbf{Q}\ket\psi,
 \end{equation}
where \(\Pi_{\mathbf{P}}\) and \(\Pi_{\mathbf{Q}}\) are the projectors onto the subspace in which all parties get the same result when measuring \(\mathbf{P}\) and \(\mathbf{Q}\), respectively. \(\Pi_{\mathbf{P}}\) is given by
\begin{equation}
    \Pi_\mathbf{P}=\left(\frac{\mathds{1}+\mathbf{P}_{AB}}2\right)\left(\frac{\mathds{1}+\mathbf{P}_{BC}}2\right)=\frac14(\mathds{1}+T_{\mathbf{P}}),\quad T_{\mathbf{P}}=\mathbf{P}_{AB}+\mathbf{P}_{BC}+\mathbf{P}_{{AC}},
\end{equation}
and similarly, \(\Pi_{\mathbf{Q}}\) is given by 
\begin{equation}
    \Pi_\mathbf{Q}=\left(\frac{\mathds{1}+\mathbf{Q}_{AB}}2\right)\left(\frac{\mathds{1}+\mathbf{Q}_{BC}}2\right)=\frac14(\mathds{1}+T_{\mathbf{Q}}),\quad T_{\mathbf{Q}}=\mathbf{Q}_{AB}+\mathbf{Q}_{BC}+\mathbf{Q}_{{AC}},
\end{equation}
where \(\mathbf{P}_{XY}\) and \(\mathbf{Q}_{XY}\) follow the shorthand of \eqref{eq:PAB}.
Substituting these expressions in \eqref{eq:probcon}, we get
 \begin{equation}\label{eq:M}
     \text{Prob(Consensus)}=\frac14+\frac18\bra\psi M\ket\psi\le\frac14+\frac{\lambda_\text{max}}8,\quad M=T_{\mathbf{P}}+T_{\mathbf{Q}},
 \end{equation}
where \(\lambda_\text{max}\) is the largest eigenvalue of \(M\). Amusingly, even though \(M\) can be a very large matrix, its spectrum is contained in \(\{-2,0,2,4\}\) (see \autoref{a:M}). This gives the inequality 
\begin{equation}\label{eq:UB}
\text{Prob(Consensus)}\leq\frac34.
\end{equation}
This scenario is a special case of the magic hypercube game for \(N\geq3\). Therefore, the probability of winning when there are three or more players is bounded by
\begin{equation}
\text{Prob(Win)}\leq\text{Prob(Consensus)}\leq\frac34.
\end{equation}
Surprisingly, this means that for three or more players it is necessarily worse to use the natural generalization of the optimal quantum strategy rather than the optimal classical strategy, since its probability of winning remains \(8/9\).

\subsubsection{A lower bound on the winning probability for three or more players}

To attain a lower bound on the winning probability of a quantum strategy, let us consider the following approach. First, the players select some row from the initial \(3\times3\) magic-square. By construction, the three observables in this row commute and thus share a common eigenbasis. The players then choose some common eigenstate \(\ket\phi\) of these observables, and prepare their initial quantum state to be
\begin{equation}
\ket\psi_{1\cdots N}=\ket\phi_1\otimes\cdots\otimes\ket\phi_N.
\end{equation}
Using this state, whenever the players measure an observable corresponding to a cell in the selected row they all obtain the same eigenvalue and therefore win. Hence, in order to find the winning probability, all that remains is to determine the probability for the common cell to correspond to an observable in that row.

\medskip

From the construction given in \eqref{eq:construction}, the \(N\)-dimensional table is obtained by copying the \((N-1)\)-dimensional table three times, up to some permutation of the rows. Consequently, each observable from the initial \(3\times3\) table makkes up exactly \(1/9\) of the \(N\)-dimensional table. Thus, the probability for the common cell to correspond to an observable in the row the players have selected is \(1/3\). We can therefore conclude that using this approach, for all \(N\) the winning probability satisfies
\begin{equation}\label{eq:qlb}
\text{Prob(Win)}\geq\frac13.
\end{equation}

\section{Concluding remarks}

To conclude, we explored a particular generalization of the Mermin-Peres magic square game to an \(N\)-dimensional magic hypercube, in which we defined the notion of a row in \(N\) dimensions as a one-dimensional slice. We showed that, for this generalization, the optimal winning probability of a classical strategy is \(8/9\) for any number of players. We also generalized the magic square construction to an \(N\)-dimensional hypercube, and proved that no quantum state can guarantee consensus among three or more players. Moreover, we derived upper and lower bounds on the winning probability of the natural generalization of the optimal two-player quantum strategy, and found that for \(N\geq3\) the upper bound is strictly below the winning probability of the optimal classical strategy. Thus, while quantum strategies outperform classical strategies for two players, the situation is reversed for three or more players.

\medskip

Several caveats should be noted. First, we have considered only one generalization of the magic square game to an \(N\)-dimensional hypercube, which may not capture the properties of other possible generalizations. Second, our analysis is restricted to Pauli measurements on qubits and does not consider more general observables or qudit systems. Finally, the upper bound in \eqref{eq:UB} is not optimal, and could be potentially made tighter.

\medskip

The failure of three or more players to achieve consensus is a reflection of monogamy. This suggests a fundamental limitation on quantum advantage in multipartite games where consensus among all players is required.

\printbibliography

\begin{appendices}

\section{A two by two seed for the magic square}\label{a:demagic}

Any valid magic square can be constructed from a \(2\times2\) block of Pauli operators:
\[
\renewcommand{\arraystretch}{1.6}
\begin{array}{|c|c|}
\hline
\mathbf{A} & \mathbf{B} \\
\hline \mathbf{C} & \mathbf{D} \\
\hline
\end{array}
\renewcommand{\arraystretch}{1}
\]
which satisfy the relations
\begin{equation}\label{eq:seedcon}
\com{\mathbf{A}}{\mathbf{B}}=\com{\mathbf{C}}{\mathbf{D}}=\com{\mathbf{A}}{\mathbf{C}}=\com{\mathbf{B}}{\mathbf{D}}=0,\quad
\acom{\mathbf{A}}{\mathbf{D}}=\acom{\mathbf{B}}{\mathbf{C}}=0.
\end{equation}
We shall refer to this block as the \emph{seed}.

\medskip

To see why, let us try to extend a given seed to a magic square. Using the product constraint over the rows and columns, together with the fact that Pauli operators square to the identity, we obtain that necessarily:
\[
\renewcommand{\arraystretch}{1.6}
\begin{array}{|c|c|c|}
\hline
\cellcolor{yellow}\mathbf{A} & \cellcolor{yellow}\mathbf{B} & \mathbf{AB} \\
\hline
\cellcolor{yellow}\mathbf{C} & \cellcolor{yellow}\mathbf{D} & \mathbf{CD} \\
\hline
\mathbf{-AC} & \mathbf{-BD} & ? \\
\hline
\end{array}
\renewcommand{\arraystretch}{1}
\]
Now, for the two topmost rows and two leftmost columns to satisfy the commutation constraint, the operators \(\mathbf{A}\), \(\mathbf{B}\), \(\mathbf{C}\) and \(\mathbf{D}\) must obey the commutation relations in \eqref{eq:seedcon}. In addition, the bottom-right cell is overdetermined by the product constraints. Hence, consistency requires
\begin{equation}\label{eq:consistency}
-\mathbf{ABCD}=\mathbf{ACBD}\Longleftrightarrow \mathbf{A}\acom{\mathbf B}{\mathbf C}\mathbf{D}=0\Longleftrightarrow\acom{\mathbf B}{\mathbf C}=0.
\end{equation}
Using these results, we get that in order for the rightmost column in the magic square to satisfy the commutation constraint, we must demand that
\begin{equation}
\com{\mathbf{AB}}{\mathbf{CD}}=
\mathbf{A}\acom{\mathbf{B}}{\mathbf{C}}\mathbf{D}-\mathbf{C}\acom{\mathbf{A}}{\mathbf{D}}\mathbf{B}-\mathbf{AC}\com{\mathbf{B}}{\mathbf{D}}-\com{\mathbf{A}}{\mathbf{C}}\mathbf{DB}=
-\mathbf{C}\acom{\mathbf{A}}{\mathbf{D}}\mathbf{B}=0\Longleftrightarrow\acom{\mathbf A}{\mathbf D}=0.
\end{equation}
This also ensures that the bottommost row in the magic square satisfies the commutation constraint, since
\begin{equation}
\com{\mathbf{AC}}{\mathbf{BD}}=
\mathbf{A}\acom{\mathbf{C}}{\mathbf{B}}\mathbf{D}-\mathbf{B}\acom{\mathbf{A}}{\mathbf{D}}\mathbf{C}-\mathbf{AB}\com{\mathbf{C}}{\mathbf{D}}-\com{\mathbf{A}}{\mathbf{B}}\mathbf{DC}=0.
\end{equation}
We thus obtain that if the seed satisfies the constraints in \eqref{eq:seedcon}, we can indeed construct from it a valid magic square.

\medskip

Let us note that no assumptions beyond the Pauli nature of \(\mathbf A\), \(\mathbf B\), \(\mathbf C\) and \(\mathbf D\) were made in this derivation, and that this construction is reversible. Meaning, every magic square composed of Pauli operators can be obtained from a seed satisfying \eqref{eq:seedcon}, and therefore must contain at least two pairs of anti-commuting observables.

\section{Stabilizers for the magic square game}\label{a:singlet}

To find the stabilizers which satisfy \eqref{eq:CC}, let us denote \(\mathbf{P}=P\otimes Q\), with  \(P,Q\in\{X,Y,Z,\mathds{1}\}\). Substituting this notation in \eqref{eq:CC}, we get
\begin{equation}
(P \otimes Q)_A \otimes (P \otimes Q)_B\ket{\psi}_{AB} = \ket{\psi}_{AB},
\end{equation}
where we also added subscripts to the operators to clarify the subsystem in which they act. Since any one of the cells in the table can be selected to be the common cell, this equation must hold for all nine cells. To solve this set of equations, let us assume that \(\ket\psi_{AB}\) is of the form
\begin{equation}\label{eq:AN}
\ket\psi_{AB}=\ket{\phi_1}_{AB}\otimes\ket{\phi_2}_{AB},
\end{equation}
where \(\ket{\phi_1}_{AB}\) is the two-qubit state of Alice's and Bob's first qubits, and \(\ket{\phi_2}_{AB}\) is the two-qubit state of their second qubits. Substituting \eqref{eq:AN} in \eqref{eq:CC} and rearranging, we get
\begin{equation}
\left(P_A \otimes P_B\ket{\phi_1}_{AB}\right)\otimes 
\left(Q_A \otimes Q_B\ket{\phi_2}_{AB}\right)
=\ket{\phi_1}_{AB}\otimes\ket{\phi_2}_{AB}.
\end{equation}
This implies that necessarily
\begin{equation}\label{eq:SE}
P_A\otimes P_B \ket{\phi_1}_{AB}=p\ket{\phi_1}_{AB}, \quad 
Q_A\otimes Q_B \ket{\phi_2}_{AB}=q\ket{\phi_2}_{AB}; \quad pq=1.
\end{equation}
Since \(P\) and \(Q\) are Pauli operators, we can derive that
\begin{equation}
\ket{\phi_1}_{AB},\ket{\phi_2}_{AB}\in\mathcal{B},
\end{equation}
where \(\mathcal{B}\) is some subset of the Bell states which depends on the exact table chosen. For the table given in \autoref{fig:quantum2}, \(P\) and \(Q\) run over all possible combinations of \(X\), \(Y\) and \(Z\). Thus, \(\ket{\phi_1}_{AB}\) and \(\ket{\phi_2}_{AB}\) must have the same eigenvalue for \(X_A \otimes X_B\), \(Y_A \otimes Y_B\) and \(Z_A \otimes Z_B\). The only Bell state that satisfies this condition is the singlet state
\begin{equation}
\ket{\Psi^-}_{AB}=\frac{\ket{0}_A\otimes \ket{1}_B-\ket{1}_A\otimes \ket{0}_B}{\sqrt 2}.
\end{equation}
Meaning,
\begin{equation}
\ket{\psi}_{AB} = \ket{\Psi^-}_{AB}\otimes \ket{\Psi^-}_{AB}.
\end{equation}

\section{The spectrum of M}\label{a:M}

To determine the constraints on the spectrum of \(M\) as defined in \eqref{eq:M}, let us compute its square, which is given by
\begin{equation}\label{eq:M2}
M^2=(T_{\mathbf{P}}+T_{\mathbf{Q}})^2=T_{\mathbf{P}}^2+T_{\mathbf{Q}}^2+\acom{T_{\mathbf{P}}}{T_{\mathbf{Q}}}.
\end{equation}
Starting with \(T_{\mathbf{P}}^2\), since its terms commute and square to the identity, we obtain
\begin{equation}
T_{\mathbf{P}}^2=(\mathbf{P}_{AB}+\mathbf{P}_{BC}+\mathbf{P}_{AC})^2=3\mathds{1}+2(\mathbf{P}_{AB}\mathbf{P}_{BC}+\mathbf{P}_{BC}\mathbf{P}_{AC}+\mathbf{P}_{AC}\mathbf{P}_{AB}).
\end{equation}
Let us notice that for the first Pauli product, we have
\begin{equation}\label{eq:Pcycle}
\mathbf{P}_{AB}\mathbf{P}_{BC}=(\mathbf{P}\otimes\mathbf{P}\otimes\mathds{1})(\mathds{1}\otimes\mathbf{P}\otimes\mathbf{P})=\mathbf{P}\otimes\mathds{1}\otimes\mathbf{P}=\mathbf{P}_{AC},
\end{equation}
and similarly for the other two products. Hence, we get
\begin{equation}\label{eq:TPsquare}
T_{\mathbf{P}}^2=3\mathds{1}+2T_{\mathbf{P}}.
\end{equation}
By the same argument, for \(T_{\mathbf{Q}}^2\) we get
\begin{equation}\label{eq:TQsquare}
T_{\mathbf{Q}}^2=3\mathds{1}+2T_{\mathbf{Q}}.
\end{equation}
Moving on to \(\acom{T_{\mathbf{P}}}{T_{\mathbf{Q}}}\), since \(\mathbf{P}\) and \(\mathbf{Q}\) anti-commute, we can derive that
\begin{equation}
\acom{T_{\mathbf{P}}}{T_{\mathbf{Q}}}=\acom{\mathbf{P}_{AB}+\mathbf{P}_{BC}+\mathbf{P}_{AC}}{\mathbf{Q}_{AB}+\mathbf{Q}_{BC}+\mathbf{Q}_{AC}}=2(\mathbf{P}_{AB}\mathbf{Q}_{AB}+\mathbf{P}_{BC}\mathbf{Q}_{BC}+\mathbf{P}_{AC}\mathbf{Q}_{AC}).
\end{equation}
This means that overall,
\begin{equation}\label{eq:M2relation}
M^2=6\mathds{1}+2M+2A,\quad A=\mathbf{P}_{AB}\mathbf{Q}_{AB}+\mathbf{P}_{BC}\mathbf{Q}_{BC}+\mathbf{P}_{AC}\mathbf{Q}_{AC}.
\end{equation}
We can now reduce the spectral problem for \(M\) to that of \(A\). Let us assume \(\ket\psi\) is an eigenstate of \(M\) with eigenvalue \(m\), from \eqref{eq:M2relation} we thus have
\begin{equation}\label{eq:Mspec}
m^2\ket\psi=(6+2m+2A)\ket\psi.
\end{equation}
Meaning, \(\ket\psi\) must also be an eigenstate of \(A\), with an eigenvalue which we will denote as \(a\). It remains to determine the constraints on the spectrum of \(A\), which we will also do by computing its square. Since the terms in \(A\) commute and square to the identity, we obtain
\begin{equation}
A^2=3\mathds{1}+2(\mathbf{P}_{AB}\mathbf{Q}_{AB}\mathbf{P}_{BC}\mathbf{Q}_{BC}+\mathbf{P}_{BC}\mathbf{Q}_{BC}\mathbf{P}_{AC}\mathbf{Q}_{AC}+\mathbf{P}_{AC}\mathbf{Q}_{AC}\mathbf{P}_{AB}\mathbf{Q}_{AB}).
\end{equation}
Let us notice that for the first Pauli product, we have
\begin{equation}
\mathbf{P}_{AB}\mathbf{Q}_{AB}\mathbf{P}_{BC}\mathbf{Q}_{BC}=-\mathbf{P}_{AB}\mathbf{P}_{BC}\mathbf{Q}_{AB}\mathbf{Q}_{BC}=-\mathbf{P}_{AC}\mathbf{Q}_{AC},
\end{equation}
and similarly for the other two products. Hence, we get
\begin{equation}\label{eq:Asquare}
A^2=3\mathds{1}-2A.
\end{equation}
Applying \(\ket\psi\) to this equation then yields
\begin{equation}
a^2\ket\psi=(3-2a)\ket\psi\Longrightarrow a\in\{-3,1\}.
\end{equation}
Therefore, from \eqref{eq:Mspec}, the eigenvalues of \(M\) must satisfy either one of the following equations
\begin{equation}
m^2-2m=0,\quad m^2-2m-8=0.
\end{equation}
This in turn gives us that the spectrum of \(M\) is contained in \(\{-2,0,2,4\}\).

\end{appendices}

\end{document}